\documentclass[epj,nopacs]{svjour}
\usepackage{graphics}
\newcommand{\sss}{\scriptscriptstyle}
\begin{document}
\title{Theoretical progress for the associated production of a Higgs
  boson with heavy quarks at hadron colliders}
\author{S.~Dawson\inst{1} \and C.~B.~Jackson\inst{2} \and
  L.~H.~Orr\inst{3} \and \underline{L.~Reina}\inst{2}\thanks{Talk
    presented by L.Reina.} \and D.~Wackeroth\inst{4}
}
\institute{Physics Department, Brookhaven National Laboratory, Upton,
  NY 11973-5000, USA \and Physics Department, Florida State
  University, Tallahassee, FL 32306-4350, USA \and Department of
  Physics and Astronomy, University of Rochester, Rochester, NY 14627,
  USA \and Department of Physics, SUNY at Buffalo, Buffalo, NY
  14260-1500, USA}
\date{Received: date/ Revised version: date}
\abstract{The production of a Higgs boson in association with a pair
  of $t\bar{t}$ or $b\bar{b}$ quarks plays a very important role at
  both the Tevatron and the Large Hadron Collider. The theoretical
  prediction of the corresponding cross sections has been improved by
  including the complete next-to-leading order QCD corrections. After
  a brief introduction, we review the results obtained for both the
  Tevatron and the Large Hadron Collider.
%
%\PACS{} % end of PACS codes
} %end of abstract
\titlerunning{Theoretical progress for associated production of a
  Higgs boson with heavy quarks at hadron colliders}
\authorrunning{L.~Reina et al.}  \maketitle
\section{Introduction}
\label{sec:intro}
The discovery and study of one or more Higgs bosons is among the most
important goals of present and future colliders. In this context, the
production of a Higgs boson with a pair of top or bottom quark and
antiquark is important both for the discovery of a Higgs boson and for
the study of the Higgs Yukawa couplings to quarks.

Observing $p\bar{p}\to t\bar{t}h$ at the Tevatron
($\sqrt{s}\!=\!2$~TeV) will require very high
luminosity~\cite{Goldstein:2000bp} and will probably be beyond the
machine capabilities. On the other hand, if $M_{h}\!\le\!130$~GeV,
$pp\to t\bar{t}h$ is an important discovery channel for a SM-like
Higgs boson at the LHC
($\sqrt{s}\!=\!14$~TeV)~\cite{atlas:1999,Beneke:2000hk,Drollinger:2001ym}.
Given the statistics expected at the LHC, $pp\to t\bar{t}h$, with
$h\to b\bar{b},\tau^+\tau^-,W^+W^-,\gamma\gamma$ will also be
instrumental to the determination of the couplings of a discovered
Higgs boson, and offer a unique handle on the top quark Yukawa
coupling~\cite{Zeppenfeld:2000td,Zeppenfeld:2002ng,Belyaev:2002ua,Maltoni:2002jr,Duehrssen:2003at}.

The associated production of a Higgs boson with a pair of $b\bar{b}$
quarks has a very small cross section in the SM, and can therefore be
used to test the hypothesis of enhanced bottom quark Yukawa couplings
which is common to many extensions of the SM, such as the MSSM for
large values of $\tan\beta$.  Both the Tevatron and the LHC will be
able to search for evidence of an enhanced $b\bar{b}h$ production,
in both inclusive and exclusive measurements. Detecting two bottom
quarks in the final state identifies uniquely the Higgs coupling
responsible for the enhanced cross section and drastically reduces the
background. This is the case we will consider in the following.

In view of their phenomenological relevance, a lot of effort has been
recently invested in improving the stability of the theoretical
predictions for the hadronic total cross sections for $p\bar{p},pp\to
t\bar{t}h$ and $p\bar{p},pp\to
b\bar{b}h$~\cite{Beenakker:2001rj,Reina:2001sf,Reina:2001bc,Beenakker:2002nc,Dawson:2002tg,Dawson:2003zu,Dittmaier:2003ej,Dawson:2003pl}.
In this proceeding we will present the results of our calculation of
the NLO cross section for both the inclusive $p\bar{p},pp\to
t\bar{t}h$
\cite{Reina:2001sf,Reina:2001bc,Dawson:2002tg,Dawson:2003zu} and the
exclusive $p\bar{p},pp\to b\bar{b}h$ cross sections
\cite{Dawson:2003pl}, where $h$ denotes the SM Higgs boson and, in the
case of $b\bar{b}h$, also the scalar Higgs bosons of the MSSM.  In
both cases the NLO cross sections have a drastically reduced
renormalization and factorization scale dependence, of the order of
15-20\% as opposed to the 100\% uncertainty of the LO cross sections.
This leads to increased confidence in prediction based on this
results.

The calculation of the NLO corrections to the hadronic processes
$p\bar{p},pp\to t\bar{t}h$ and $p\bar{p},pp\to b\bar{b}h$ presents
challenging technical difficulties, ranging from virtual pentagon
diagrams with several massive internal and external particles to real
gluon and quark emission in the presence of infrared singularities.
We refer to \cite{Reina:2001bc,Dawson:2003zu} for a complete
discussion of all technical details.

\boldmath
\section{Results for $t\bar{t}h$ production}
\label{sec:tth}
\begin{figure}[tbh]
\begin{center}
\resizebox{0.45\textwidth}{!}{%
  \includegraphics{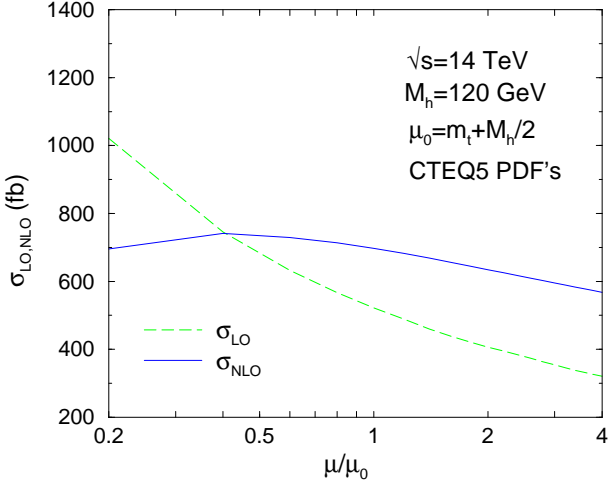}
}
\caption[]{Dependence of $\sigma_{\sss LO,NLO}(pp\to t\bar{t}h)$ on 
  the renormalization/factorization scale $\mu$, at $\sqrt{s_{\sss
      H}}\!=\!14$~TeV, for $M_h\!=\!120$ GeV.}
\label{fg:tth_mudep_lhc}
\end{center}
\end{figure}

\begin{figure}[tbh]
\begin{center}
\resizebox{0.45\textwidth}{!}{%
  \includegraphics{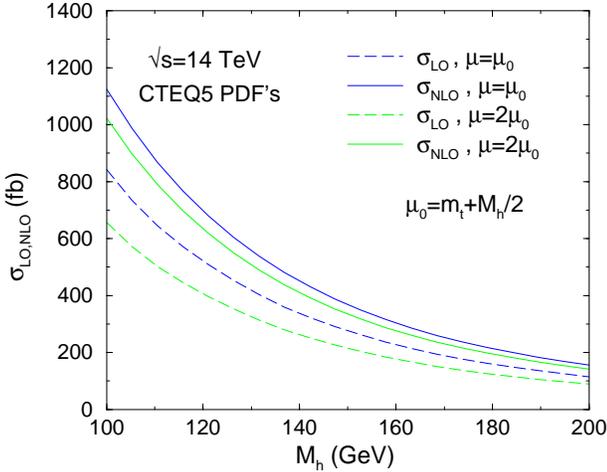}
}
\caption[]{$\sigma_{\sss NLO}(pp\to t\bar{t}h)$ and $\sigma_{\sss
    LO}(pp\to t\bar{t}h)$ as functions of $M_h$, at $\sqrt{s_{\sss
      H}}\!=\!14$~TeV, for $\mu\!=m_t+M_h/2$ and $\mu\!=\!2m_t+M_h$.}
\label{fg:tth_mhdep_lhc}
\end{center}
\end{figure}

\unboldmath The impact of NLO QCD corrections on the tree level cross
section for $pp\to t\bar{t}h$ production (LHC) in the SM is
illustrated in Figs.~\ref{fg:tth_mudep_lhc} and
\ref{fg:tth_mhdep_lhc}. Similar results for the case of $p\bar{p}\to
t\bar{t}h$ production (Tevatron) can be found in
Ref.~\cite{Reina:2001sf,Reina:2001bc}.  Results for $\sigma_{\sss
  NLO}$ ($\sigma_{\sss LO}$) are obtained using the 2-loop (1-loop)
evolution of $\alpha_s(\mu)$ and CTEQ5M (CTEQ5L) parton distribution
functions \cite{Lai:1999wy}, with $\alpha_s^{\sss
  NLO}(M_Z)\!=\!0.118$. The top quark mass is renormalized in the OS
scheme and its pole mass is fixed at $m_t\!=\!174$~GeV.
Fig.~\ref{fg:tth_mudep_lhc} illustrates the
renormalization/factorization scale dependence of $\sigma_{\sss LO}$
and $\sigma_{\sss NLO}$ at the LHC. The NLO cross section shows a
drastic reduction of the scale dependence with respect to the lowest
order prediction.  Fig.~\ref{fg:tth_mhdep_lhc} complements this
information by illustrating the dependence of the LO and NLO cross
sections on the Higgs boson mass at the LHC. For scales
$\mu\!\ge\!0.4\mu_0$ ($\mu_0\!=\!m_t+M_h/2$) the NLO corrections
enhance the cross section. We estimate the remaining theoretical
uncertainty on the NLO result to be of the order of 15-20\%, due to
the leftover $\mu$-dependence, the error from the PDFs, and the
error on the top quark pole mass $m_t$.

\boldmath
\section{Results for $b\bar{b}h$ production}
\label{sec:bbh}
\unboldmath 
We evaluate the fully exclusive cross section for $b\bar{b}h$
production by requiring that the transverse momentum of both final
state bottom and anti-bottom quarks be larger than 20~GeV
($p_T^b\!>\!20$~GeV), and that their pseudorapidity satisfy the
condition $|\eta_b|\!<\!2$ for the Tevatron and $|\eta_b|\!<\!2.5$ for
the LHC.  This corresponds to an experiment measuring the Higgs decay
products along with two high $p_T$ bottom quark jets. In order to
better simulate the detector response, the final state gluon and the
bottom/anti-bottom quarks are treated as distinct particles only if
the separation in the azimuthal angle-pseudorapidity plane is $\Delta
R\!>\!0.4$.  For smaller values of $\Delta R$, the four momentum
vectors of the two particles are combined into an effective
bottom/anti-bottom quark momentum four-vector.

The set-up used for the NLO (LO) calculation of the $b\bar{b}h$ cross
section is the same as for $t\bar{t}h$ (see Sec.~\ref{sec:tth}). In
the $b\bar{b}h$ case, however, we have also investigated the
dependence of the NLO result on the choice of the renormalization
scheme for the bottom quark Yukawa coupling.  The strong scale
dependence of the $\overline{MS}$ bottom quark mass
($\overline{m}_b(\mu)$) plays a special role in the perturbative
evaluation of the $b\bar{b}h$ production cross section since it enters
in the overall bottom quark Yukawa coupling.  The same is not true for
$t\bar{t}h$ production since the $\overline{MS}$ top quark mass has
only a very mild scale dependence.  The bottom quark pole mass is
taken to be $m_b\!=\!4.6$~GeV. In the OS scheme the bottom quark
Yukawa coupling is calculated as $g_{b\bar{b}h}\!=\!m_b/v$, while in
the $\overline{MS}$ scheme as
$g_{b\bar{b}h}(\mu)\!=\!\overline{m}_b(\mu)/v$, where we use the
2-loop (1-loop) $\overline{MS}$ bottom quark mass for the NLO (LO)
cross section respectively.
\begin{figure}[bth]
\begin{center}
\resizebox{0.45\textwidth}{!}{%
  \includegraphics{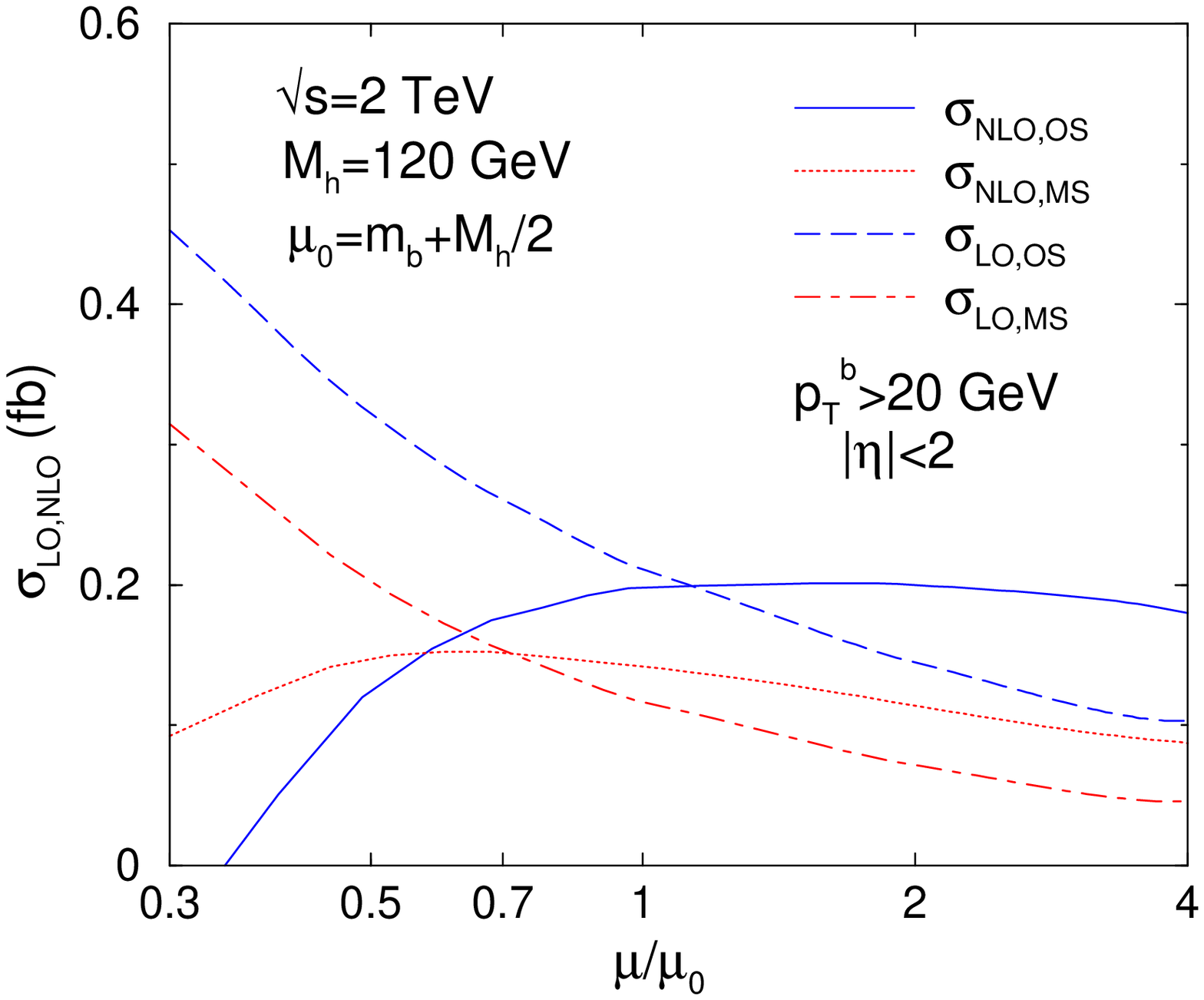}
}
\resizebox{0.45\textwidth}{!}{%
  \includegraphics{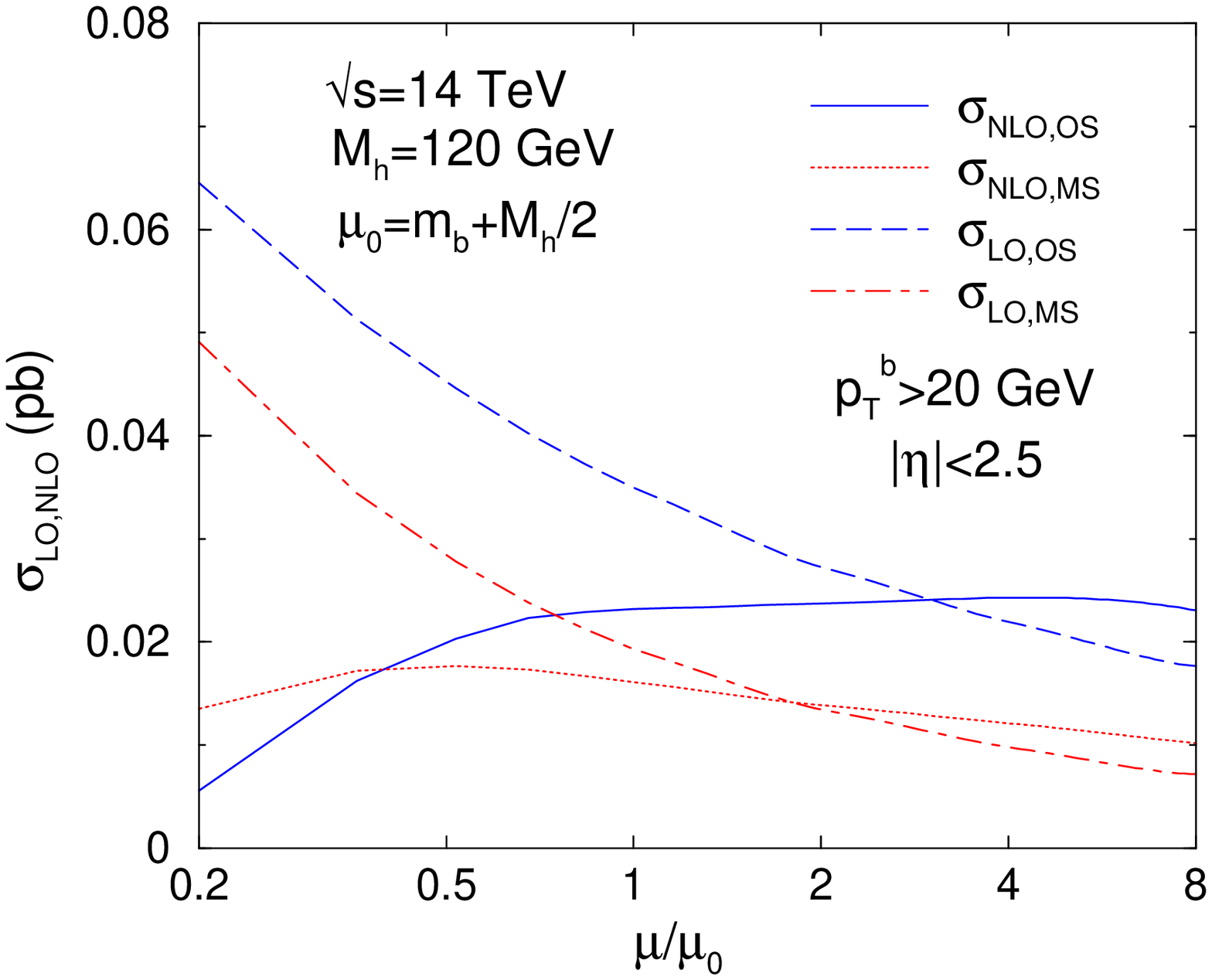}
}
\caption[]{$\sigma_{\sss NLO}$ and $\sigma_{\sss LO}$ for $p\bar{p}\to
  b\bar{b}h$ at $\sqrt{s}\!=\!2$~TeV (top) and for $pp\to b\bar{b}h$
  at $\sqrt{s}\!=\!14$~TeV (bottom) as a function of the
  renormalization/factorization scale $\mu$, for $M_h=120$~GeV.}
\label{fg:bbh_mu_dep}
\end{center}
\end{figure}

The impact of NLO QCD corrections on the tree level cross section for
$b\bar{b}h$ exclusive production in the SM is summarized in
Fig.~\ref{fg:bbh_mu_dep} for both the Tevatron and the LHC. In both
the $OS$ and the $\overline{MS}$ schemes the stability of the cross
section is greatly improved at NLO, and the corresponding theoretical
uncertainty reduced to 15-20\%. The $\overline{MS}$ results seem to
have overall a better perturbative behavior, although the variation of
the NLO cross section about its point of least sensitivity to the
renormalization/factorization scale is almost the same when one uses
the $OS$ or the $\overline{MS}$ schemes for the bottom Yukawa
coupling.  This indicates that the running of the Yukawa coupling is
not the only important factor to determine the overall perturbative
stability of the cross section. The difference between the $OS$ and
the $\overline{MS}$ results at their plateau values should probably be
interpreted as an additional theoretical
uncertainty~\cite{Dawson:2003pl}.

\begin{figure}[t]
\begin{center}
\resizebox{0.45\textwidth}{!}{%
  \includegraphics{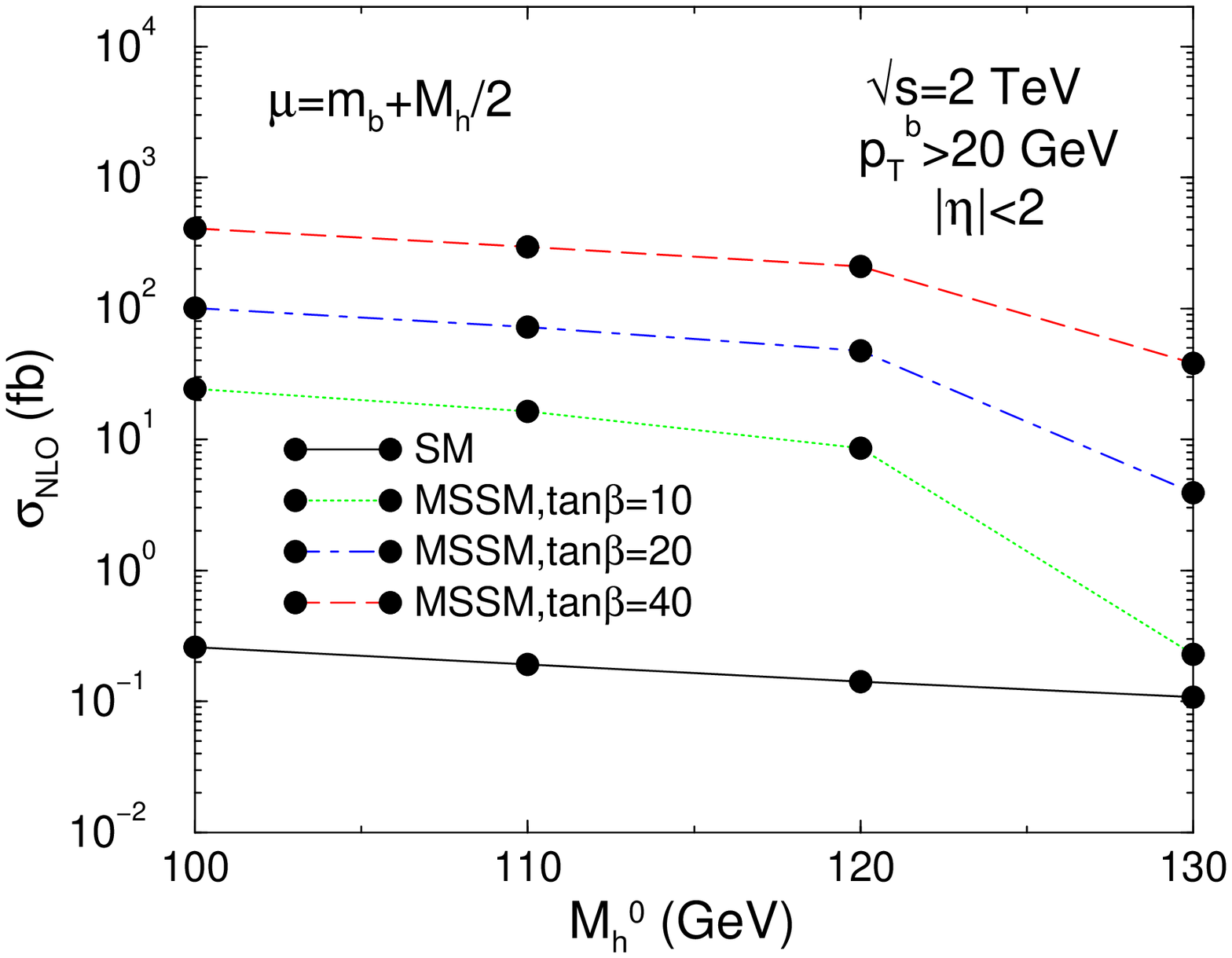}
}
\resizebox{0.45\textwidth}{!}{%
  \includegraphics{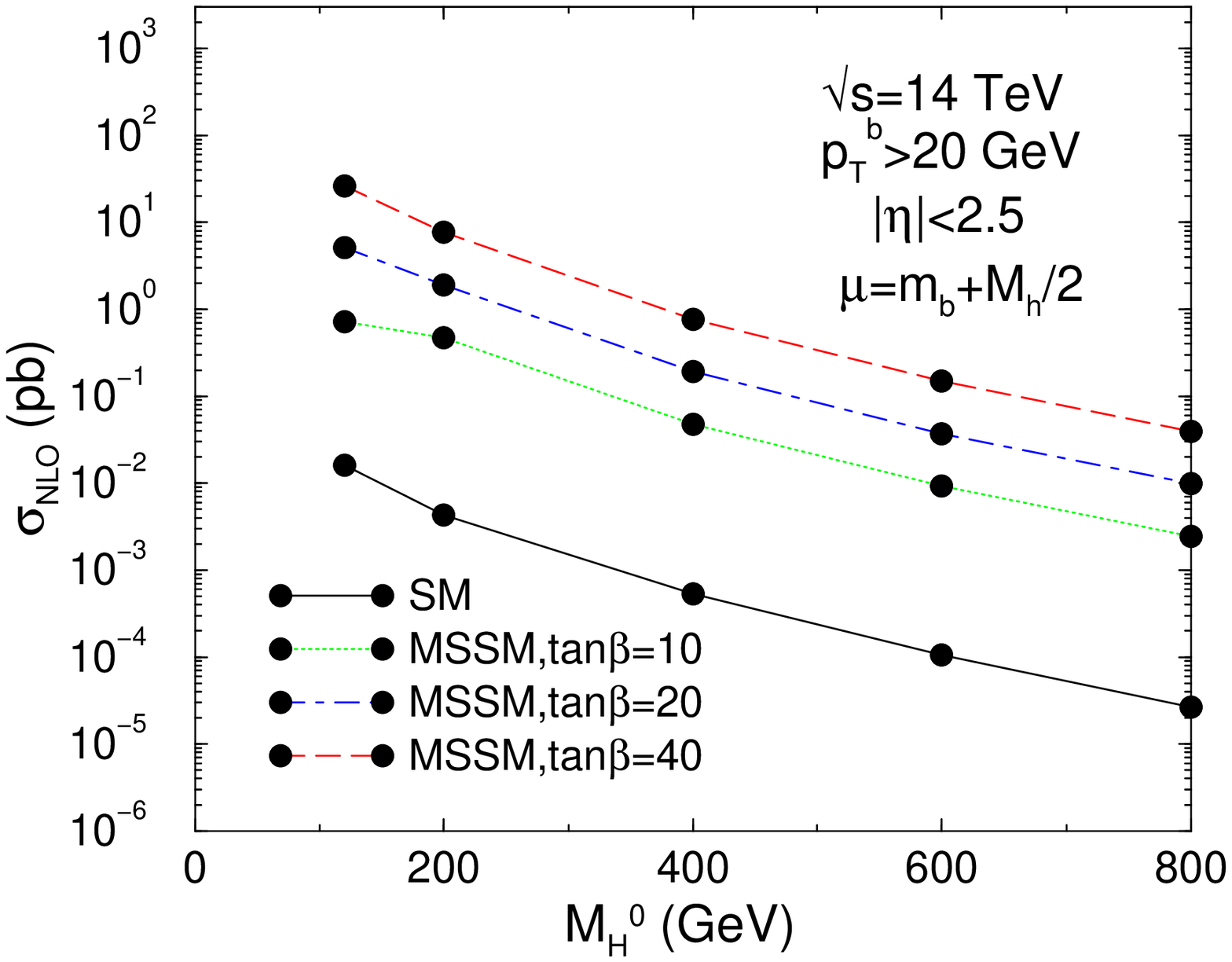}
}
\caption[]{ $\sigma_{\sss NLO,MS}$ 
  for $p\bar{p}\to b\bar{b}h$ production at $\sqrt{s}\!=\!2$~TeV (top)
  and $pp\to b\bar{b}h$ production at $\sqrt{s}\!=\!14$~TeV (bottom)
  in the SM and in the MSSM with $\tan\beta\!=\!10,20$, and $40$.}
\label{fg:bbh_mh_dep}
\end{center}
\end{figure}

Finally, in Fig.~\ref{fg:bbh_mh_dep} we illustrate the dependence of
the exclusive cross section, at the Tevatron and at the LHC, on the
Higgs boson mass, both in the SM and in some scenarios of the MSSM,
corresponding to $\tan\beta\!=\!10,20$, and $40$. For the Tevatron we
consider the case of the light MSSM scalar Higgs boson($h^0$) while
for the LHC we consider the case of the heavy MSSM scalar Higgs boson
($H^0$). We see that the rate for $b\bar{b}h$ production can be
significantly enhanced in a supersymmetric model with large values of
$\tan\beta$, and makes $b\bar{b}h$ a very important mode for discovery
of new physics at both the Tevatron and the LHC.

\section*{Acknowledgments}
The work of S.D. (C.J./L.R., L.H.O.)  is supported in part by the U.S.
Department of Energy under grant DE-AC02-98CH10886 (DE-FG02-97ER41022,
DE-FG-02-91ER40685). The work of D.W. is supported in part by the
National Science Foundation under grant No.~PHY-0244875.
\bibliography{../../qcd2003/proceedings/qcd03}

\begin{thebibliography}{10}
\providecommand*{\bibinfo}[2]{#2}
\providecommand*{\eprint}[1]{#1}
\providecommand*{\url}[1]{#1}
\bibitem{Goldstein:2000bp}
\bibinfo{author}{J.~Goldstein} \emph{et~al.}, \bibinfo{journal}{Phys. Rev.
  Lett.} \bibinfo{volume}{\textbf{86}}, \bibinfo{pages}{1694}
  (\bibinfo{date}{2001}), \eprint{hep-ph/0006311}.
\bibitem{atlas:1999}
\bibinfo{author}{{ATLAS Collaboration}} (\bibinfo{date}{1999}), {Technical
  Design Report, Vol. II}, \eprint{{CERN/LHCC/99-15}}.
\bibitem{Beneke:2000hk}
\bibinfo{author}{M.~Beneke} \emph{et~al.} (\bibinfo{date}{2000}),
  \eprint{hep-ph/0003033}.
\bibitem{Drollinger:2001ym}
\bibinfo{author}{V.~Drollinger}, \bibinfo{author}{T.~Muller}, and
  \bibinfo{author}{D.~Denegri} (\bibinfo{date}{2001}), \eprint{hep-ph/0111312}.
\bibitem{Zeppenfeld:2000td}
\bibinfo{author}{D.~Zeppenfeld}, \bibinfo{author}{R.~Kinnunen},
  \bibinfo{author}{A.~Nikitenko}, and \bibinfo{author}{E.~Richter-Was},
  \bibinfo{journal}{Phys. Rev.} \bibinfo{volume}{\textbf{D62}},
  \bibinfo{pages}{013009} (\bibinfo{date}{2000}), \eprint{hep-ph/0002036}.
\bibitem{Zeppenfeld:2002ng}
\bibinfo{author}{D.~Zeppenfeld} (\bibinfo{date}{2002}),
  \eprint{hep-ph/0203123}.
\bibitem{Belyaev:2002ua}
\bibinfo{author}{A.~Belyaev} and \bibinfo{author}{L.~Reina},
  \bibinfo{journal}{JHEP} \bibinfo{volume}{\textbf{08}}, \bibinfo{pages}{041}
  (\bibinfo{date}{2002}), \eprint{hep-ph/0205270}.
\bibitem{Maltoni:2002jr}
\bibinfo{author}{F.~Maltoni}, \bibinfo{author}{D.~Rainwater}, and
  \bibinfo{author}{S.~Willenbrock}, \bibinfo{journal}{Phys. Rev.}
  \bibinfo{volume}{\textbf{D66}}, \bibinfo{pages}{034022}
  (\bibinfo{date}{2002}), \eprint{hep-ph/0202205}.
\bibitem{Duehrssen:2003at}
\bibinfo{author}{M.~D{\"u}rssen} (\bibinfo{date}{2003}), {ATL/PHYS-2003-30}.
\bibitem{Beenakker:2001rj}
\bibinfo{author}{W.~Beenakker}, \bibinfo{author}{S.~Dittmaier},
  \bibinfo{author}{M.~Kr{\"a}mer}, \bibinfo{author}{B.~Pl{\"u}mper},
  \bibinfo{author}{M.~Spira}, and \bibinfo{author}{P.~Zerwas},
  \bibinfo{journal}{Phys. Rev. Lett.} \bibinfo{volume}{\textbf{87}},
  \bibinfo{pages}{201805} (\bibinfo{date}{2001}), \eprint{hep-ph/0107081}.
\bibitem{Reina:2001sf}
\bibinfo{author}{L.~Reina} and \bibinfo{author}{S.~Dawson},
  \bibinfo{journal}{Phys. Rev. Lett.} \bibinfo{volume}{\textbf{87}},
  \bibinfo{pages}{201804} (\bibinfo{date}{2001}), \eprint{hep-ph/0107101}.
\bibitem{Reina:2001bc}
\bibinfo{author}{L.~Reina}, \bibinfo{author}{S.~Dawson}, and
  \bibinfo{author}{D.~Wackeroth}, \bibinfo{journal}{Phys. Rev.}
  \bibinfo{volume}{\textbf{D65}}, \bibinfo{pages}{053017}
  (\bibinfo{date}{2002}), \eprint{hep-ph/0109066}.
\bibitem{Beenakker:2002nc}
\bibinfo{author}{W.~Beenakker}, \bibinfo{author}{S.~Dittmaier},
  \bibinfo{author}{M.~Kr{\"a}mer}, \bibinfo{author}{B.~Pl{\"u}mper},
  \bibinfo{author}{M.~Spira}, and \bibinfo{author}{P.~Zerwas},
  \bibinfo{journal}{Nucl. Phys.} \bibinfo{volume}{\textbf{B653}},
  \bibinfo{pages}{151} (\bibinfo{date}{2003}), \eprint{hep-ph/0211352}.
\bibitem{Dawson:2002tg}
\bibinfo{author}{S.~Dawson}, \bibinfo{author}{L.~H. Orr},
  \bibinfo{author}{L.~Reina}, and \bibinfo{author}{D.~Wackeroth},
  \bibinfo{journal}{Phys. Rev.} \bibinfo{volume}{\textbf{D67}},
  \bibinfo{pages}{071503} (\bibinfo{date}{2003}), \eprint{hep-ph/0211438}.
\bibitem{Dawson:2003zu}
\bibinfo{author}{S.~Dawson}, \bibinfo{author}{C.~Jackson},
  \bibinfo{author}{L.~H. Orr}, \bibinfo{author}{L.~Reina}, and
  \bibinfo{author}{D.~Wackeroth}, \bibinfo{journal}{Phys. Rev.}
  \bibinfo{volume}{\textbf{D68}}, \bibinfo{pages}{034022}
  (\bibinfo{date}{2003}), \eprint{hep-ph/0305087}.
\bibitem{Dittmaier:2003ej}
\bibinfo{author}{S.~Dittmaier}, \bibinfo{author}{M.~Kr{\"a}mer}, and
  \bibinfo{author}{M.~Spira} (\bibinfo{date}{2003}), \eprint{hep-ph/0309204}.
\bibitem{Dawson:2003pl}
\bibinfo{author}{S.~Dawson}, \bibinfo{author}{C.~Jackson},
  \bibinfo{author}{L.~Reina}, and \bibinfo{author}{D.~Wackeroth}
  (\bibinfo{date}{2003}), \eprint{hep-ph/0311067}.
\bibitem{Lai:1999wy}
\bibinfo{author}{H.~L. Lai} \emph{et~al.} (\bibinfo{collaboration}{CTEQ}),
  \bibinfo{journal}{Eur. Phys. J.} \bibinfo{volume}{\textbf{C12}},
  \bibinfo{pages}{375} (\bibinfo{date}{2000}), \eprint{hep-ph/9903282}.

\end{thebibliography}
\end{document}